# The effect of MgO insertion into the interstitial spaces of $BaY_2O_4$ on the upconversion response of $Er^{3+}$:$Yb^{3+}$ at illumination with 976 nm excitation light

**Liviu Dudaș**

National University of Science and Technology Politehnica Bucharest, Faculty of Chemical Engineering and Biotechnologies, 1-7 Gheorghe Polizu Street, 011061 Bucharest, Romania; liviu_dudas@yahoo.com (L.D.),

**Abstract**: The efficiencies and form of the upconversion spectra of rare earth ions embedded in crystalline matrices strongly depend upon the phonon dispersion of these, since the inter-level energy transition probabilities are influenced by the phonon energies. The aim of this study is to show a novel method of influencing the phonon energy values of crystalline $BaY_2O_4$ and, by this, to finely tune the spectra of any rare earth ion as a dopant of it. We used $Er^{3+}$ sensitized by $Yb^{3+}$ as dopants, but the same method can be generalized for other kind of rare earth ions. As such, we synthesized $BaY_2O_4$ and $BaY_2O_4$:MgO ($BaY_2MgO_5$) ceramics having as activator-sensitizer pairs the $Er^{3+}$ 1% ion and $Yb^{3+}$ 2% relative to the substituted $Y^{3+}$. Firstly, we synthesized the mix of precursor metallic oxides by the citrate-gel method, pressed these into pellets, and sintered them to obtain doped crystalline ceramic samples. The ceramic samples were illuminated with 976 nm from a laser diode, and the resulting upconversion spectra of $Er^{3+}$ were measured. $BaY_2O_4$ and $BaY_2O_4$:MgO ($BaY_2MgO_5$) have almost the same XRD signature, showing that the crystal unit cells have the same parameters but they differ in the phononic energies due to the presence of the additional MgO. Comparing the results of these two extreme cases— $BaY_2O_4$ vs. $BaY_2O_4$:MgO ($BaY_2MgO_5$)—reveals that the presence of the MgO modifies the upconversion response. So, this study fulfills two goals: First, it reveals the structure of the undocumented $BaY_2O_4$:MgO ($BaY_2MgO_5$) and shows how it is based on the crystal structure of $BaY_2O_4$ with $Mg^{2+}$ ions occupying the unit cell interstitions. Secondly, it suggests a method of tuning the phononic energies of $BaY_2O_4$ and, as a consequence, the upconversion response of $Er^{3+}$.

Keywords: $BaY_2O_4$, $BaY_2O_4$:MgO, $BaY_2MgO_5$:$Er^{3+}$,$Yb^{3+}$; upconversion; citrate-gel method; photoluminescence

## 1. Introduction

Rare earths (RE) ions have the peculiarity that they have the electrons in the 4f shell closer to the nucleus than the 5d 6s valence electrons [1,2,3]. This isolation allows for very distinct and large separated energy levels, which lead to sharp emission lines in a large range of wavelengths of interest.

Among the many types of crystalline materials studied for the upconversion properties of $Er^{3+}$ ions as dopants, either alone or sensitized by $Yb^{3+}$, there are those with the general formula $BaLn_2ZnO_5$ (with Ln=Y,Gd), which are used for efficient luminescent phosphors or high-sensitivity thermometry [4,5].

There can be seen that the unit cells of $BaGd_2ZnO_5$ (BGZO) and $BaY_2ZnO_5$ (BYZO) have the same structure, orthorhombic, with $Y^{3+}$ ions replacing $Gd^{3+}$ ones [6]. Because $Y^{3+}$ ions have a smaller ionic radius than $Gd^{3+}$ (0.96 Å vs. 1.00 Å [7]), the unit cell for BYZO has slightly smaller dimensions, having a volume of 515.866 Å$^3$, while BGZO has 529.955 Å$^3$ [6].

We studied the upconversion emission of $Er^{3+}$:$Yb^{3+}$ as an activator-sensitizer ionic pair doping these two kinds of crystalline matrices under illumination with 976 nm from a laser diode. We remarked that the intenstity of the upconversion emission, in the case of BYZO, is at least three times stronger than that in the case of BGZO, for the same intensity of the excitation.

This effect (all the other structural parameters being almost the same) should be due to the fact that the phonon frequencies in the case of BYZO, which have the same ion placements in the unit cell, with $Y^{3+}$ ions replacing the $Gd^{3+}$ ones, are different than those in BGZO.

As shown in [8], the lesser the energy of the phonons, the lower the probability of multiphonon (MP) assisted transitions between the energy levels of the rare-earth ions. In order to increase the efficiency of the UC process, the MP transitions must be controlled, and this should be done by controlling the frequency of the phonons.

The frequencies of the phonons, $\omega$, are governed by the elastic constants of the structure and the masses of the ions. In crystals, normal modes are classified by a wave-vector $q$. Phonon frequencies, $\omega(q)$, and displacement patterns of these normal modes, $U_s^\alpha(q)$, are determined by the following secular equations where $\tilde{C}_{st}^{\alpha\beta}$ are the Fourier transforms from the spatial domain of the inter-atomic force constants $C_{IJ}^{\alpha\beta} \equiv \frac{\partial^2 E(R)}{\partial R_I^\alpha \partial R_J^\beta}$ [9]:

$$\sum_{t,\beta}\left(\tilde{C}_{st}^{\alpha\beta}(q) - M_s\omega^2(q)\delta_{st}\delta_{\alpha\beta}\right)U_t^\beta(q) = 0$$

Due to the transversal nature of the electromagnetic radiation, the phonon modes involved are the transversal optical (TO) ones.

So, in order to further tweak the phonon energies, we tried to replace the $Zn^{2+}$ ion with one that has similar characteristics: +2 valence, no coloring (due to completely filled shells of the ion), and similar ionic radius for V coordination [7]. From all atomic species, only Mg has these traits, having 2+ stable valence, no coloring, and similar crystal and ionic radii as Zn.

We tried to prepare the $BaY_2O_4$:MgO ($BaY_2MgO_5$) (BYMO) ceramic by the same procedure by which we synthesized BYZO, expecting that a unit cell similar to BYZO would be obtained, but, after several tries, we only found that the XRD patterns were those for $BaY_2O_4$, with no patterns of MgO. The only explanation was that Mg ions were occupying the interstitial spaces that one can see in **Figure 1**(A) with data taken from [6] legacy-3952.

Also, we manually inserted in the cif file Mg and O ions in the interstitions (**Figure 1**(B), and the simulated XRD was similar to the one obtained for the samples, both BYO and BYMO.

A research on information regarding $BaY_2O_4$:MgO ($BaY_2MgO_5$), neither [6] nor another specific information source gives any results. This shows that this material is new and could find some future application.

Magnesium oxide ICDD card [01-071-11776] (periclase) is shown in **Figure 3** to indicate that MgO didn't segregate when inserted in BYMO. Peaks from the shown card are almost undetectable in the BYMO-measured diffractogram.

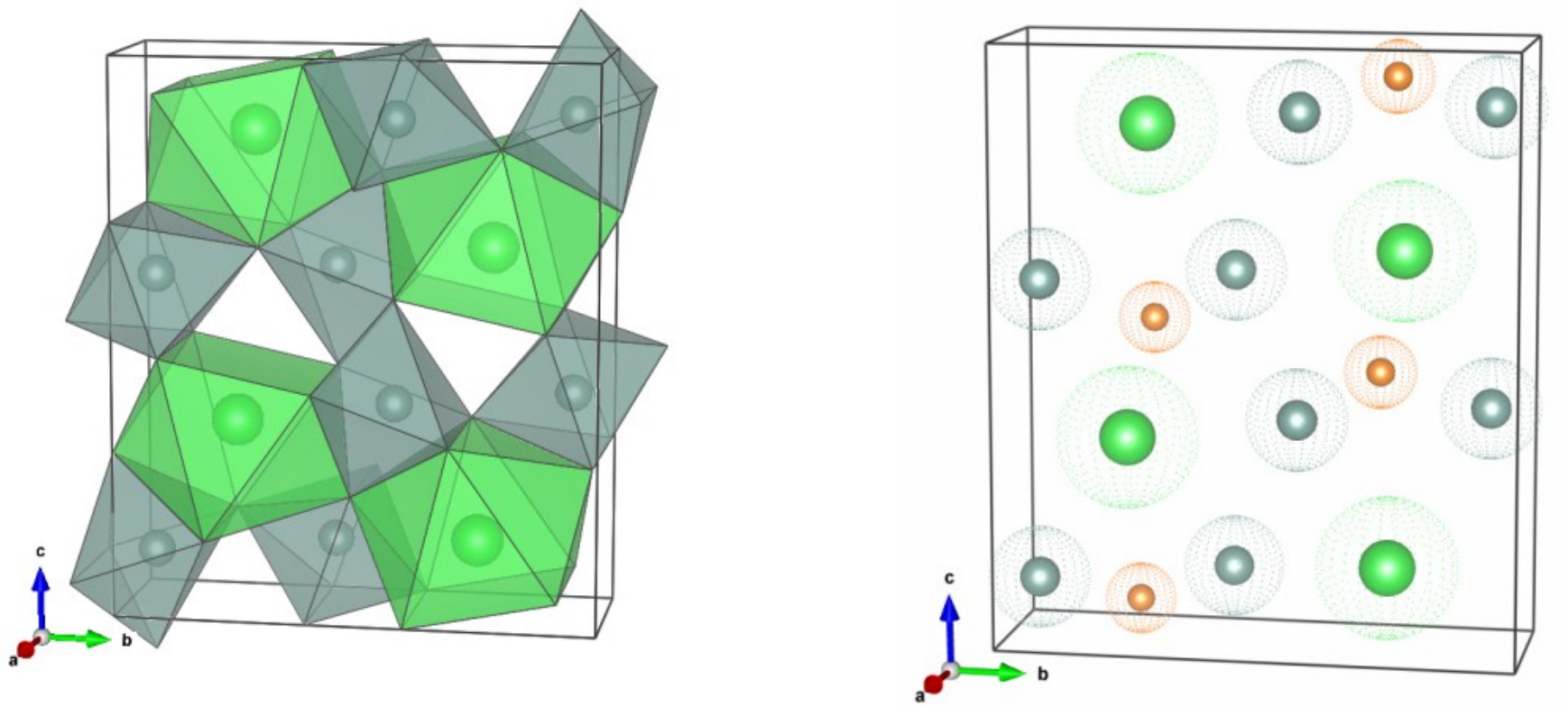

(A) (B)

**Figure 1**. (A): Surrounding polyhedra of $Y^{3+}$ and $Ba^{2+}$ in $BaY_2O_4$. (B) $Mg^{2+}$ ions (with orange) manually inserted in the interstitial spaces of $BaY_2O_4$ in order to form $BaY_2O_4$:MgO ($BaY_2MgO_5$). $Ba^{2+}$ ions are shown with green, $Y^{3+}$ ions are shown with gray.

## 2. Materials and Methods

### *2.1. Materials and the method used*

The preparation of $BaY_2O_4$:MgO ($BaY_2MgO_5$) and $BaY_2O_4$ ceramics was done through the citrate-gel method, in which, firstly, the metallic oxides were prepared and secondly, the powder oxides were compacted into pellets and thermally treated at 1200 °C for 3 h.

The starting reagents were research grade barium (II) acetate crystalline form (BA), tri-magnesium dicitrate nonahydrate (MC), yttrium (III) nitrate hexahydrate (YN), erbium tri-nitrate pentahydrate (EN), ytterbium tri-nitrate pentahydrate (YbN), citric acid (CA), ammonia solution 25% (AS), and ethylenedinitrilotetraacetic acid (EDTA), whose source was Sigma Aldrich (part of Merck Group, Darmstadt, Germany).

The chosen concentrations for the dopant ions were 1% for $Er^{3+}$ and 2% for $Yb^{3+}$ relative to the substituted yttrium atoms.

For the rest of the article, the samples will be labeled BYO 1-2 for $BaY_2O_4$: Er(1%): Yb(2%) and BYMO 1-2 for $BaY_2O_4$:MgO ($BaY_2MgO_5$): Er(1%): Yb(2%), respectively.

These relative and absolute $Er^{3+}$:$Yb^{3+}$ concentrations of the dopants were chosen because, firstly, we observed in our studies that the highest intensity upconversion signal is obtained for this ratio, and secondly, the total dopant concentration does not determine the segregation of other different crystalline phases.

### *2.2. Synthesis of oxide powders*

Appropriate quantities of the metallic nitrates and an appropriate quantity of CA (2 CA moles for each mole of metallic ions) were dissolved in EG.

Instead of MC, we initially used magnesium nitrate, which proved very difficult to work with because, when dissolved, magnesium quickly became hydroxicarbonated [18,18] and got out of solution, and the attempts to resolubilize it were very inefficient.

So we opted for MC, which is also sensitive to hydration but poses less problems than $MgNO_3$.

First, we prepared stock solutions of barium acetate (BA), yttrium nitrate (YN), erbium nitrate (EN), and ytterbium nitrate (YbN). We dissolved CA in distilled water at 50 C, 2.5 molecules of CA for every metallic ion, and mixed the solutions of YN, EN, and YbN.

Separately, we mixed the BA solution with EDTA, one molecule of EDTA for each $Ba^{2+}$ ion, because $Ba^{2+}$ is better chelated by this acid instead of CA.

We mixed these two solutions under slow magnetic stirring (100 rpm) at 50 °C and, after 5 minutes, added the MC in crystalline form and mixed until complete dissolution.

The slightest of turbidity after this procedure is an indication that the operation failed and magnesium got carbonated and precipitated out of the solution.

The reason for slow stirring is for the solution not to trap $CO_2$ and precipitate the magnesium. Under continuous stirring, we started the evaporation of water by raising the temperature to 90 °C.

During the water evaporation, we periodically monitored the pH of the solution to be between 8 and 9 and adjusted it with the help of the AS.

This pH value is the minimum necessary for maintaining the high concentration of $CA^{3-}$ for a proper chelation of the metallic ions [10].

The evaporation continued until a thick and clear gel formed. Then we started to slowly increase the temperature until 400 °C and calcinated for 3 hours until a grayish powder was obtained.

### *2.3. Obtaining of Er, Yb doped $BaY_2O_4$ and $BaY_2O_4$:MgO ($BaY_2MgO_5$) ceramics*

The obtained oxidic powders were further cleared of carbonaceous traces by calcination at 900 °C for 3 hours, after which 13 mm diameter and 1 mm thickness pellets were compressed at 100 $kN/cm^2$. The pressed pellets were sintered at 1250 °C for 8 hours.

These parameters were chosen in order for $Y_2O_3$ not to segregate from $BaY_2O_4$ because $Y_2O_3$ has the formation energy per atom of 3.980 eV while $BaY_2O_4$ has only 3.665 eV [6] (BaO has 2.831 eV).

Great attention must be paid for the humidity not to come into contact with the resulted samples because $BaY_2O_4$:MgO ($BaY_2MgO_5$) is very hygroscopic due to MgO, which hydrates and carbonates very quickly [18]. $BaY_2O_4$ is less sensitive yet it must be protected also from the surrounding atmosphere..

### *2.4. Characterisation of $BaY_2O_4$ and $BaY_2O_4$:MgO ($BaY_2MgO_5$) samples*

With the help of a Rigaku Miniflex II X-ray diffractometer using Cu K$\alpha_1$ line, (Tokyo, Japan), the XRD spectra of the ceramic pellets were measured in the $2\theta$ range of 15°–70°, step 0.01° and speed 2 °/min.

We spotwise illuminated the pellets with a laser 980 nm diode (LCU98E042Ap from Laser Components GmbH, Olching, Germany) and measured the upconversion spectra with a USB4000CG-UV-NIR spectrometer (Ocean Optics- Orlando, Florida, USA). The aquisition software was OceanView version 1.6.7.

The measurement setup and process are described in [14].

## 3. Results

### *3.1 XRD spectra of $BaY_2O_4$ and $BaY_2O_4$:MgO ($BaY_2MgO_5$) Ceramics*

The measured spectra of BYO 1-2 and BYMO 1-2 are presented in **Figures 2,3**. They are generated with the help of [11] software. Added are the simulated X-ray diffractogram for $BaY_2O_4$ with crystallographic data from material 3952 from [6] in which $Mg^{2+}$ ions were manually placed on the assumption that the voids in the structure are the only places where $Mg^{2+}$ can be inserted without generating initial structure deformations. The $O^{2-}$ ions (not shown) associated with $Mg^{2+}$ ones are aligned along the $\vec{a}$ direction of the crystal and inserted between $Mg^{2+}$ ions.

As seen, in comparation with XRDs for two types of MgO phases, this assumption was correct, the measured XRD for BYO 1-2 and BYMO 1-2 showing no traces of segregated MgO.

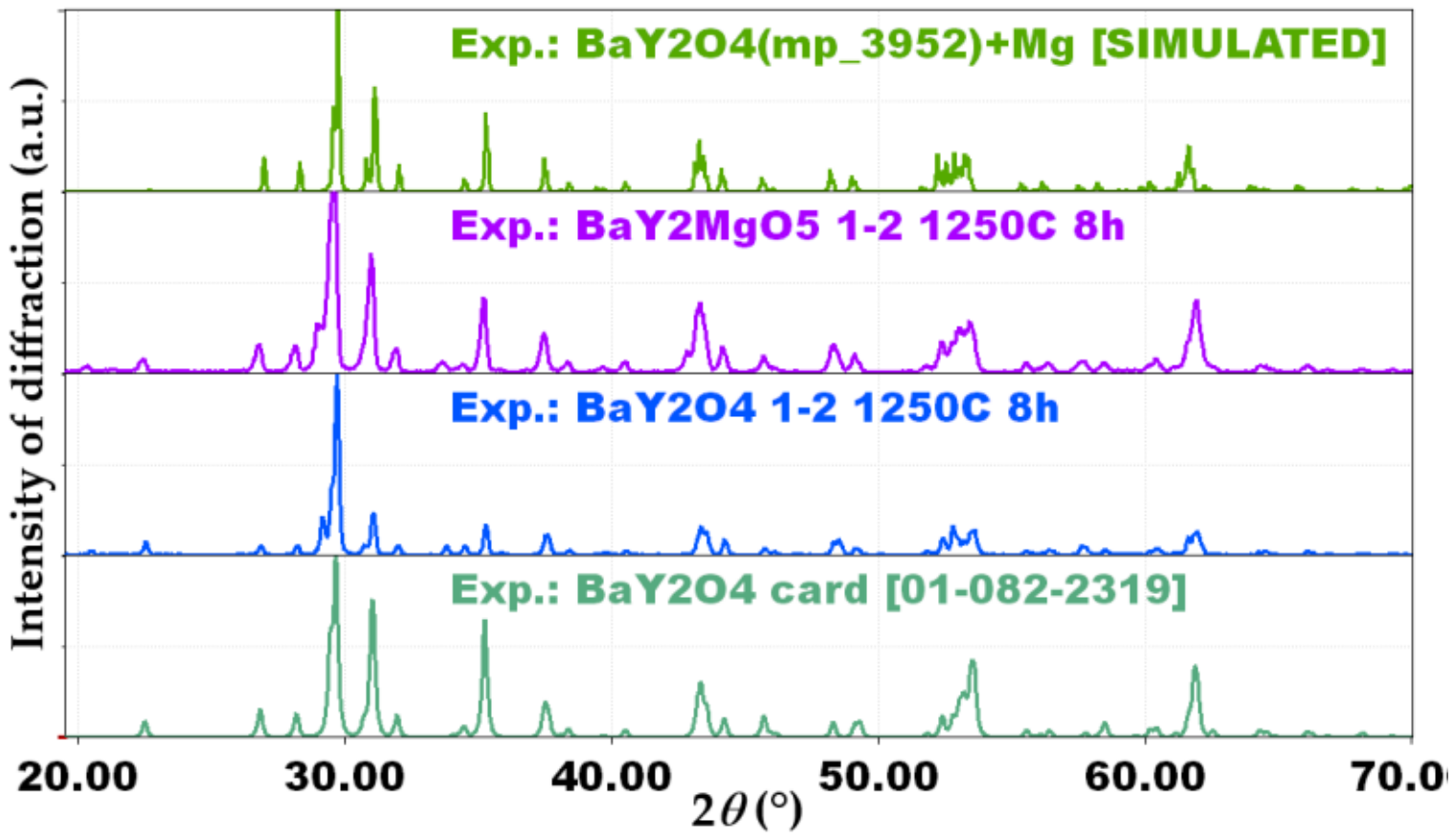

**Figure 2.** XRD patterns of, from top to bottom: (a) simulated $BaY_2O_4$ (mp_3952 from [6]) with Mg inserted into vacant places, (b), (c) measured BYMO 1-2 and BYO 1-2, (d), ICDD card [01-082-2319]. One can see the good resemblance of these, which show that the assumption that Mg enters the BYO matrix in the vacant spaces is a realistic one, verified by the experiment.

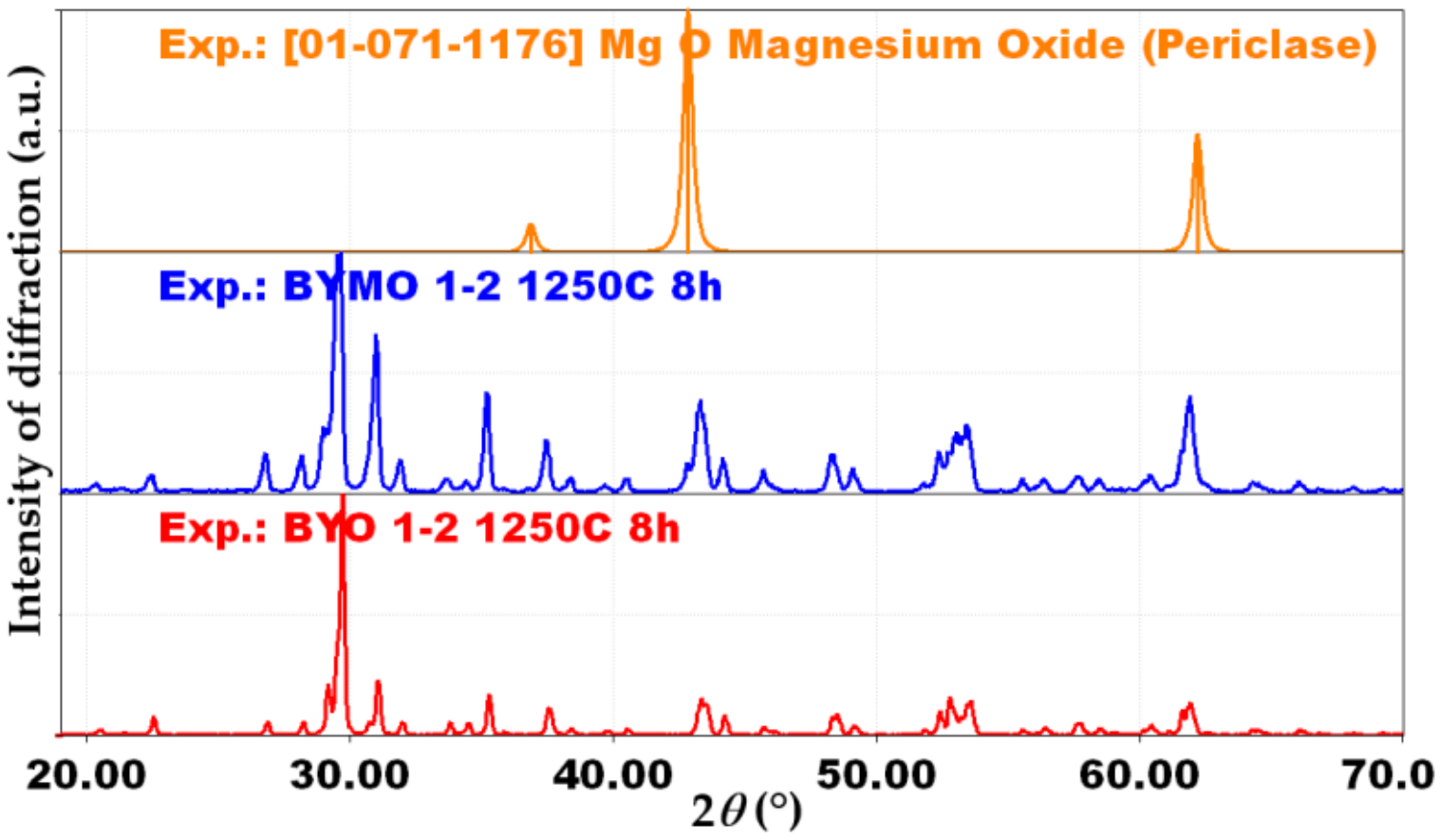

**Figure 3.** Comparison between data from ICDD card 01-71-1176 and the measured ones for BYO 1-2 and BYOM 1-2 showing that no segregated MgO phase has formed, the BYO matrix accomodating the the $Mg^{2+}$ ions very well.

### *3.2. Structure of $BaY_2O_4$ and $BaY_2O_4$:MgO ($BaY_2MgO_5$) crystal unit cell*

$BaY_2O_4$ unit cell belongs to the orthorhombic system with Pnma-62 symmetry. It has the lengths *a* = 3.489 Å, *b* = 10.523 Å, and *c* = 12.245 Å. The crystallographic data from [6] was visualised with VESTA 3 software [12] and the figure 4(C) show the coordination poyhedra of $Ba^{2+}$ and $Y^{3+}$.

Yttrium ions are VI coordinated and all have the same shape of the coordination polyhedra.

From **Figures 1** and **4**, one can observe that the matrix has voids which are large enough to accommodate $Mg^{2+}$ ions. Also, the figure 4(D) shows $Mg^{2+}$ ions, manually placed in the $BaY_2O_4$ unit cell, basing on the supposition that these are the only places where $Mg^{2+}$ ions can fit into the structure.

As the X-ray diffractograms show in the **Figures 2** and **3**, this supposition was correct, because, as seen in **Table 1**, the crystal radius of $Mg^{2+}$ is small enough to occupy the voids.

Therefore, one can conclude that this is the structure of $BaY_2O_4$:MgO ($BaY_2MgO_5$), a structure which seems to be novel, since not only the literature lacks data about it, but also [6] database doesn't contain it.

**Table 1.** Ionic and crystal radii for $Mg^{2+}$ and Zn2+

| Ion in VII coordination | Crystal radius (Å) | Ionic radius (Å) |
|---|---|---|
| $Mg^{2+}$ | 0.80 | 0.66 |
| $Zn^{2+}$ | 0.82 | 0.68 |

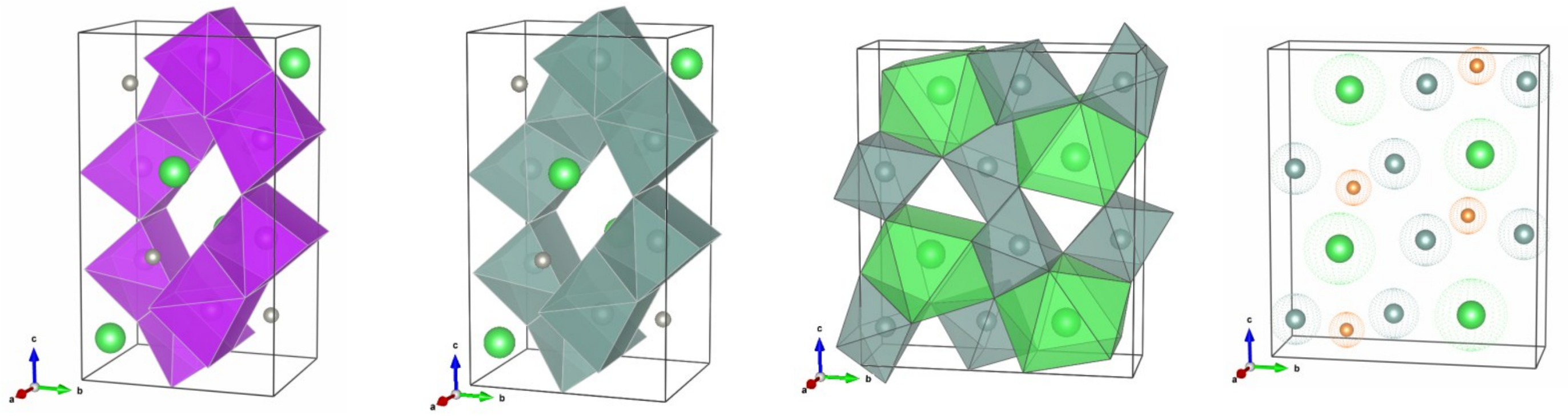

(A) (B) (C) (D)

**Figure 4** (A),(B) Orthorhombic Pnma-62 structures of $BaGd_2ZnO_5$ and $BaY_2ZnO_5$ (magenta: $Gd^{3+}$ coordination polyhedra, gray polyhedra for $Y^{3+}$; (C) Orthorhombic Pnma-62 for $BaY_2O_4$, gray polyhedra for $Y^{3+}$ and light green polyhedra for Ba2+; (D) the placement of $Mg^{2+}$ ions (orange) into the void zones of $BaY_2O_4$ from (C). The $Mg^{2+}$ ions are well accomodated, inserting no deformation into the structure.

### *3.3. Upconversion spectra*

In **Figure 5** are presented the upconversion spectra measured for BYO 1-2, BYMO 1-2 and, for peaks positions comparation, also the spectrum for $Y_2O_3$ 1-2 [13] is shown because it resemble closest those for BYO 1-2 and BYOM 1-2 with respect to green/red ratio. The $Y_2O_3$ spectrum is scaled down at 33% because the upconversion signal for $Y_2O_3$ is higher than that of BYO, for the same incident illumination power.

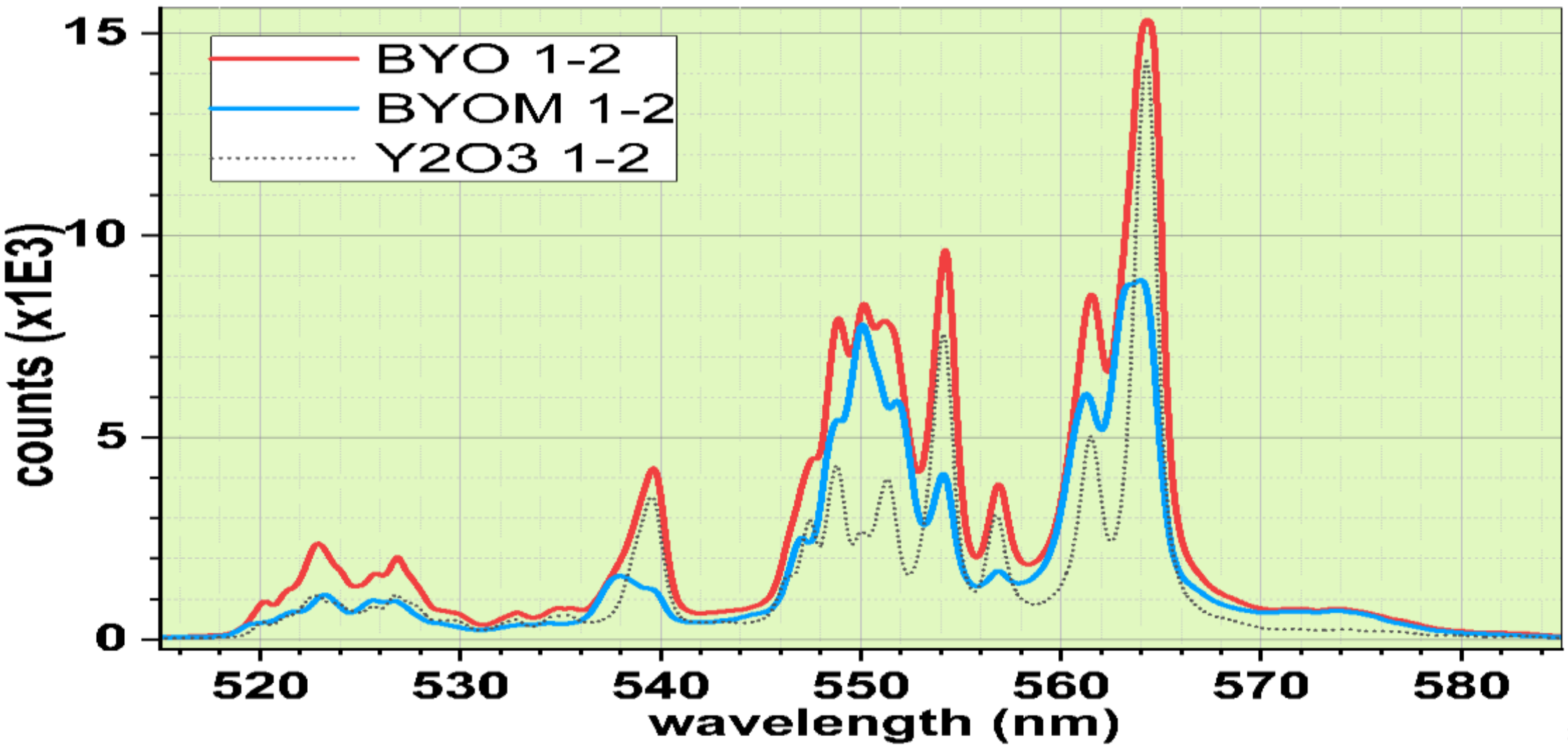

**(A)** Er$^{3+}$: $^2H_{11/2} \rightarrow ^4I_{15/2}$ Er$^{3+}$: $^4S_{3/2} \rightarrow ^4I_{15/2}$

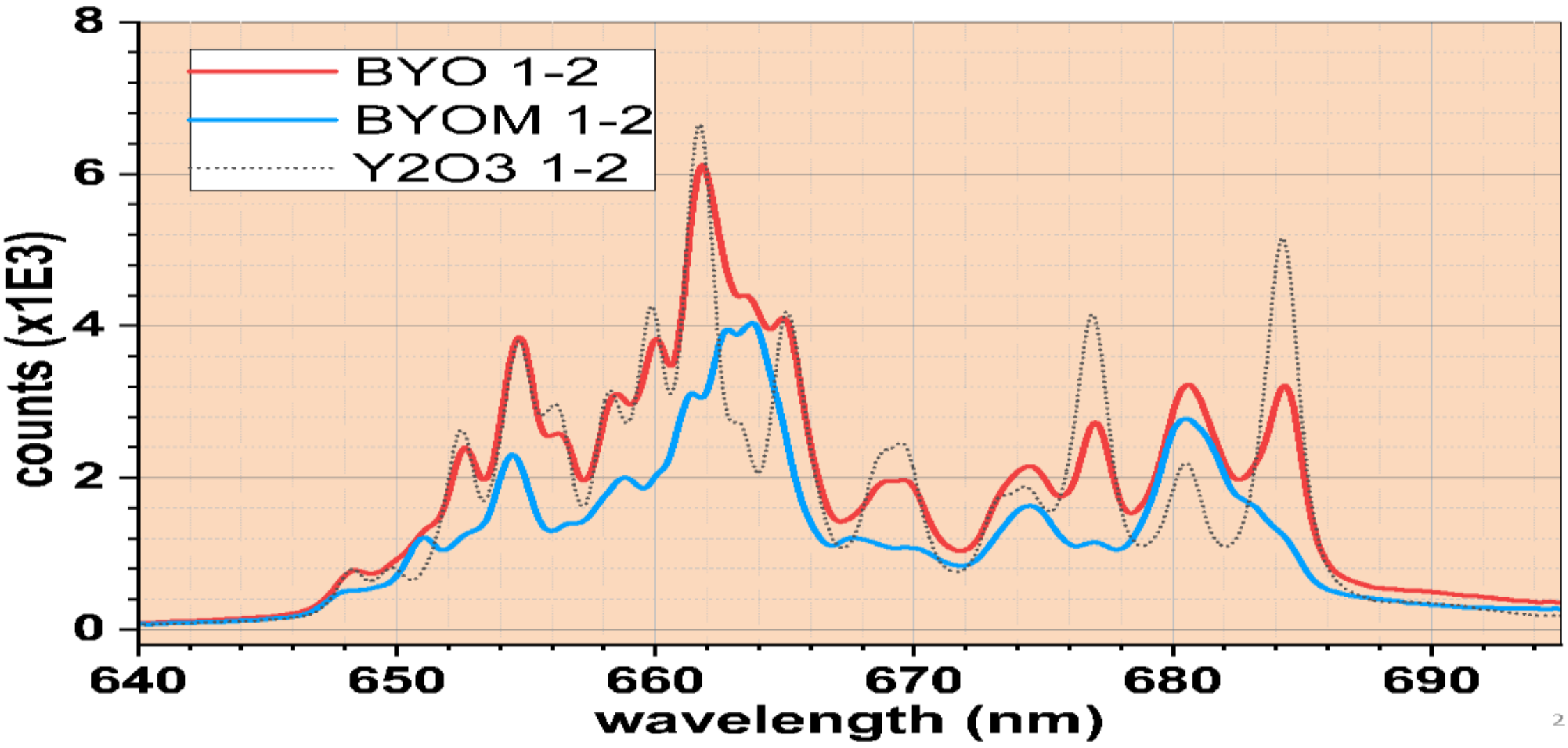

**(B)** Er$^{3+}$: $^4F_{9/2} \rightarrow ^4I_{15/2}$

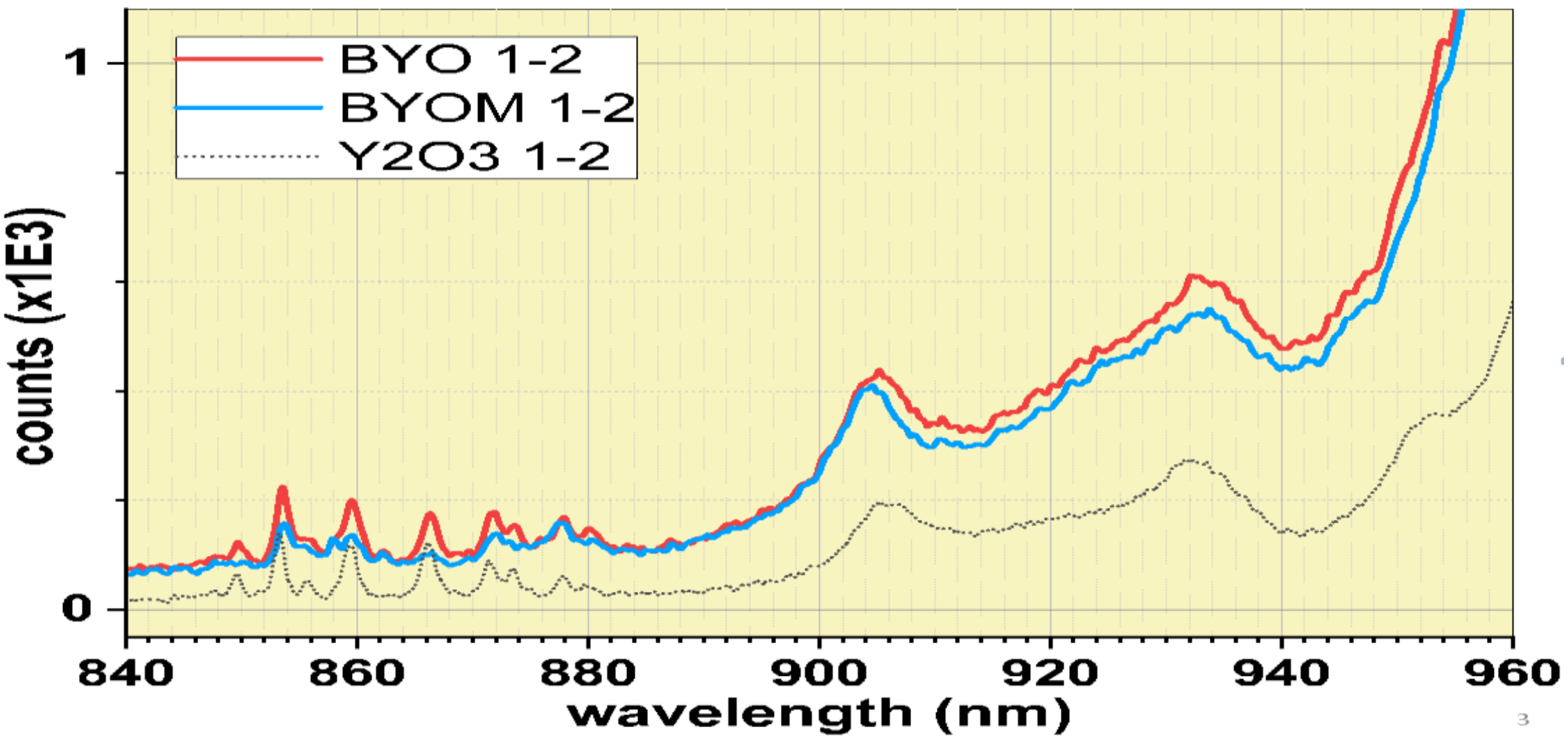

**(C)** Er$^{3+}$:$^4F_{9/2} \rightarrow ^4I_{13/2}$ (840-885 nm) Yb$^{3+}$: $^2F_{5/2} \rightarrow ^2F_{7/2}$ anti-Stokes band (900-940 nm)

**Figure 5**. (A), (B) Visible and (C) NIR upconversion spectra of BYMO 1-2 and BYO 1-2. The spectrum for $Y_2O_3$ 1-2 is also presented in order to underline the differences and also for the detection of emission peaks and grasp some idea of the strength of the crystal fields of BYO vs $Y_2O_3$, which, as it can be seen, are comparable. $Y_2O_3$ signal is scaled down at 33% in order for the graphs to be easily compared.

## 4. Discussion

### 4.1. Comments on upconversion spectra

The presence of MgO in BYO modifies the phonon dispersion of pure BYO and, as such, the upconversion response of $Er^{3+}$. One can see that the response is altered, for the same injected power, the efficiency is lower.

The $Yb^{3+}$ anti-Stokes peaks at 905 nm and 935 nm are the same in all cases, revealing from their split, that, while the phonon energies for $Y_2O_3$ and BYO and BYMO are about the same (median at 920 nm, which implies a 620 $cm^{-1}$), the strength of the crystal field in BYO and BYOM is higher (because the distance between peaks is larger)

### 4.2. Degradation

After six months, at the attempt to re-measure the spectral response of the samples, we found that they degraded, being swollen and broken.

This was because, negligently, they were not protected from the atmospheric conditions, so they hydrated and carbonated [18] following the equations:

- $MgO + CO_2 \rightarrow MgCO_3$ + 100.6 kJ/mol.
- $MgO(s) + H_2O(l) \rightarrow Mg(OH)_2(s)$ + 81.2 kJ/mol.
- $BaO + CO_2 \rightarrow BaCO_3$ + 267.2 kJ/mol.
- $BaO(s) + H_2O(l) \rightarrow Ba(OH)_2(s)$ + 78.7 kJ/mol

Nevertheless, we measured the upconversion spectra of the degraded shards and found the spectra in **Figure 6**.

The spectra are almost identical with that of YO 1-2.

The carbonation took place in the bulk and separated the metallic ions;, yet the structure remained crystalline because the sharpness of the peaks is unmodified.

This is an undoubtedly indication that the sensitization of $Er^{3+}$ by $Yb^{3+}$ is governed by their relative interdistance more than the structure of embedding.

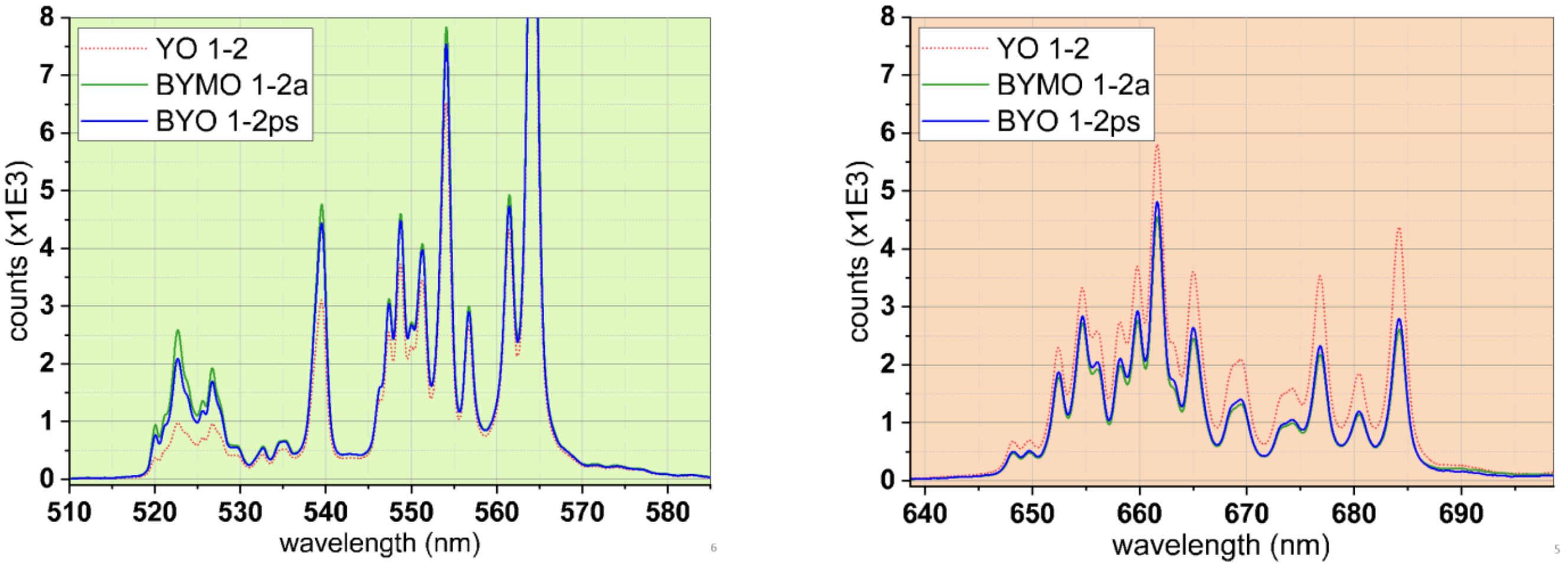

(A) (B)

**Figure 6**. Spectral response of the degraded pellets BYMO 1-2 and BYO 1-2 after carbonation and hydration, in the usual atmospheric environment, after 6 months. Their spectral response is identical with that of YO 1-2 which is shown with red dotted line, indicating that $Y_2O_3$ segregated and the other metals formed other compounds, most likely hydrated $BaCO_3$ and $MgCO_3$..

## 5. Conclusions

$BaY_2O_4$ and $BaY_2O_4$:MgO ($BaY_2MgO_5$) ceramic pellets, doped with $Er^{3+}$:$Yb^{3+}$, were fabricated by mold pressing and sintering the precursor metallic oxide mix obtained by the citrate-gel method.

The X-ray investigation showed that $BaY_2O_4$ and $BaY_2O_4$:MgO ($BaY_2MgO_5$) have almost identical diffractograms, with the $Mg^{2+}$ ions being very well accommodated into the intestitial spaces of $BaY_2O_4$.

Also, the dopant ions had a good insertion into the matrix, substituting $Y^{3+}$ ions without signs of other phases segregation or crystalline structure distortion.

To the best of our knowledge, this is the very time the compound $BaY_2O_4$:MgO ($BaY_2MgO_5$) is characterized, the literature lacking any information about it.

The samples were irradiated with an NIR laser at 976 nm, and the upconversion emission spectra of $Er^{3+}$ were measured.

The peaks in the spectra for BYMO 1-2 were less distinct than those for BYO 1-2, showing that $Mg^{2+}$ ions hinder the efficiency of the upconversion by inserting additional energy loss channels.

Yet notable were the positions of the peaks, which are the same as for $Y_2O_3$, showing that the crystal field in $BaY_2O_4$ and $BaY_2O_4$:MgO ($BaY_2MgO_5$) has the same strength as in $Y_2O_3$.

Also, the anti-Stokes band of $Yb^{3+}$ is similar in all cases of BYO, BYMO, and $Y_2O_3$, indicating an average phonon energy of 620 $cm^{-1}$ for all these oxidic ceramics.

The emission of $^4F_{9/2}$->$^4I_{13/2}$ of $Er^{3+}$ is hindered in BYMO, and $Mg^{2+}$ ions inserted in BYO are clearly the cause. The mechanism by which Mg alters the UC of Er+ in BYO will be the subject of further research, both experimental and by analyzing the phonon dispersions in BYO and BYMO through ab-initio computing techniques.

Also, the unfortunate event of degradation of the samples turned into an occasion of confirming that the sensitization of $Er^{3+}$ by $Yb^{3+}$ is governed by their relative interdistance.

A general application of these findings is that the $Er^{3+}$ UC response at 980 NIR can be fine-tuned by controlling the Mg content in the matrix of BYO.

Another application would be to detect the presence and/or concentrations of $Mg^{2+}$ ions that penetrate the BYO matrix and compare the resulted UC spectrum with that for pure BYO.

## 6. Acknowledgments and thanks

I am endebted to both Prof. D. Berger and Prof. C. Matei from the National University of Science and Technology Politehnica Bucharest, Faculty of Chemical Engineering and Biotechnologies, for allowing me to synthesize, in their laboratory, the samples that are subject to this article.

I am endebted to Prof. D. Berger for making the XRD measurements and to Prof. C. Matei for making the XRD and SEM investigations of the degraded samples.

I am endebted to Prof. C. Matei for discovering some errors after performing the SEM analysis and also for the reminder that $BaY_2O_4$:MgO is more of a solid solution or rather a mixed crystal than a stoichiometric compound, as the formula $BaY_2MgO_5$ would imply.

EOF